\documentclass[reprint,aps,pra,superscriptaddress,showpacs]{revtex4-1}
\usepackage{graphicx}
\usepackage{amsmath}
\usepackage{xcolor}
\usepackage{amsfonts}
\usepackage{braket}

\begin{document}

\title{Complex Edge-State Phase Transitions in 1D Topological Laser Arrays}

\author{Midya Parto}
\author{Steffen Wittek}
\author{Hossein Hodaei}
\affiliation{CREOL/College of Optics and Photonics, University of Central Florida, Orlando, Florida 32816, USA}
\author{Gal Harari}
\author{Miguel A. Bandres}
\affiliation{Department of Physics, Technion–Israel Institute of Technology, Haifa 32000, Israel}
\author{Jinhan Ren}
\affiliation{CREOL/College of Optics and Photonics, University of Central Florida, Orlando, Florida 32816, USA}
\author{Mikael C. Rechtsman}
\affiliation{Department of Physics, The Pennsylvania State University, University Park, Pennsylvania 16802-6300, USA}
\author{Mordechai Segev}
\affiliation{Department of Physics, Technion–Israel Institute of Technology, Haifa 32000, Israel}
\author{Demetrios N. Christodoulides}
\email{demetri@creol.ucf.edu}
\author{Mercedeh Khajavikhan}
\email{mercedeh@creol.ucf.edu}

\affiliation{CREOL/College of Optics and Photonics, University of Central Florida, Orlando, Florida 32816, USA}

\date{\today}

\begin{abstract}
% insert abstract here
We report the first observation of lasing in topological edge states in a 1D Su-Schrieffer-Heeger active array of resonators. We show that in the presence of chiral-time ($\mathcal{CT}$) symmetry, this non-Hermitian topological structure can experience complex phase transitions that alter the emission spectra as well as the ensued mode competition between edge and bulk states. The onset of these phase transitions is found to occur at the boundaries associated with the complex geometric phase- a generalized version of the Berry phase in Hermitian settings. Our experiments and theoretical analysis demonstrate that the topology of the system plays a key role in determining its operation when it lases: topologically controlled lasing.
\end{abstract}


\maketitle

In condensed matter physics, topological insulators (TI) represent new forms of matter wherein electron conduction is prohibited in the bulk, while it is allowed along the surface by means of edge states \cite{TKNN,Hasan,QSHE,Zhang_Topo}. These gapless edge states emerge whenever a TI is terminated either on vacuum or is interfaced with an ordinary insulator - a principle known as bulk-edge correspondence. This property stems from the fact that any transition between two distinct topological phases cannot be performed in a continuous fashion, but instead requires a bandgap-crossing at the interface of two materials that exhibit different topological invariants. Consequently, topological edge states are robust against local perturbations, since their characteristics are dictated by their corresponding bulk environment. This is in sharp contrast to conventional defect states that originate from imperfections, and are by nature sensitive to perturbations. It is this robustness that has incited a flurry of activities aimed to understand and harness the ramifications of topology in many and diverse fields ranging from electromagnetism \cite{Haldane_Realization,Marin_Immune,Poli_Selective} and optics \cite{Topo_Photon,Hafezi_Robust,Khanikaev,Hafezi_Imaging,Moti_Floquet,Topo_Waveguiding} to ultracold atomic gases \cite{Topo_Cold,Bloch_Zak}, mechanics \cite{Topo_Mech}, and acoustics \cite{Topo_Acoustic}. As indicated in recent studies, the introduction of topology in photonics can lead to a host of intriguing and unexpected results. These include for example unidirectional light transport, backscattered-free light propagation as well as immunity to structural imperfections \cite{Marin_Immune}. Lately, there has been a great deal of interest in studying the interplay between non-Hermiticity and topology \cite{Malzard_Topo,Topo_Dissipation,Nori_Edge}. In this context, photonics provides a versatile platform to perform such studies, since non-Hermiticity can be readily established through the introduction of optical gain and loss - an aspect that has been exploited in observing $\mathcal{PT}$-symmetric interactions and spontaneous symmetry breaking effects \cite{PT,Hodaei,Lumer_PRL}. As demonstrated in recent works, the isomorphism between the Schr\"odinger equation and the optical wave equation can be fruitfully utilized to design photonic lattice structures capable of displaying topological phenomena akin to those encountered in condensed matter physics \cite{Haldane_Realization,Hafezi_Robust,Moti_Floquet,Khanikaev,Topo_Waveguiding,Poli_Selective,Hafezi_Imaging}. An archetypical example of one-dimensional discrete lattices that is known to be topologically non-trivial (thus allowing edge modes), is that described by the Su-Schrieffer-Heeger (SSH) model \cite{SSH}. Thus far, this class of SSH structures has been employed to experimentally probe topological phase transitions \cite{Rudner_Levitov,Moti_Transition} and to demonstrate $\mathcal{PT}$-symmetric topologically protected bound states in passive systems (not involving gain) such as fused silica coupled waveguide arrays \cite{Moti_nmat}. Yet, in spite of the intense activity in this area, much remains unexplored and several fundamental questions still remain unanswered. For example, can non-Hermiticity and/or nonlinearity impede or assist topological edge states? In that case, how do topological attributes depend on the gain/loss levels?

Here we address the above questions by theoretically and experimentally investigating topological aspects in one-dimensional SSH laser arrays - structures that are both nonlinear and highly non-Hermitian. We show that the conventional chiral $\mathcal{C}$-symmetry associated with the passive SSH system no longer persists in the presence of non-Hermiticity. Instead, by judiciously engineering the gain and loss profile in the SSH laser array in a way that respects $\mathcal{PT}$-symmetry, the ensuing Hamiltonian now possesses $\mathcal{CT}$-symmetry, a necessary ingredient in this case for robust topologically protected lasing edge states. Here the complex band structure is used to predict the behavior of the active SSH structure for different levels of non-Hermiticity. Based on this analysis, we identify three different phases that depend on the gain levels involved and the coupling strengths. We find that this rich nonlinear and non-Hermitian system displays a broad range of behaviors, starting from single edge-mode lasing and eventually ending into multimode emission within the bulk of the array. Moreover, the complex Berry phase associated with the corresponding non-Hermitian SSH Hamiltonian is found to reflect these phase transitions. The observed intensity mode profiles and spectra emitted by this topological laser arrangement are in good agreement with theoretical predictions that account for carrier dynamics, saturable gain, and laser mode competition. In fact, these latter processes, play a crucial role in stabilizing the lasing edge-modes in such topological arrangements.

For our study, we fabricate an active SSH array consisting of 16 identical coupled microring resonators fabricated on InGaAsP quantum wells. The gain-medium consists of six vertically stacked quantum wells, each composed of a $10 nm$ thick well $(In_{x=0.56}Ga_{1-x}As_{y=0.93}P_{1-y})$ sandwiched between two $20 nm$ thick barrier layers $(In_{x=0.74}Ga_{1-x}As_{y=0.57}P_{1-y})$, with an overall height of $200 nm$, which is capped with a $10 nm$ thick layer of InP, as depicted in Fig. \ref{Fig1} (a). The coupling strengths in this SSH structure alternate between $\kappa_1 \approx 8 \times 10^{10} s^{-1}$ and $\kappa_2 \approx 14 \times 10^{10} s^{-1}$, as obtained when the distance between successive rings is $200 nm$ and $150 nm$, respectively. The resulting topological lattice shown in Fig. \ref{Fig1} (b) involves a nontrivial termination, capable of supporting edge modes. Figure \ref{Fig1} (c) shows a microscope image of a fabricated SSH structure (eight unit cells), with each ring being weakly coupled to a waveguide that happens to be equipped with two out-coupling gratings - necessary to interrogate the array. Within the tight-binding formalism, the dynamics in this SSH configuration can be described by the following Hamiltonian:
\begin{figure}[h]
\includegraphics[scale=0.4]{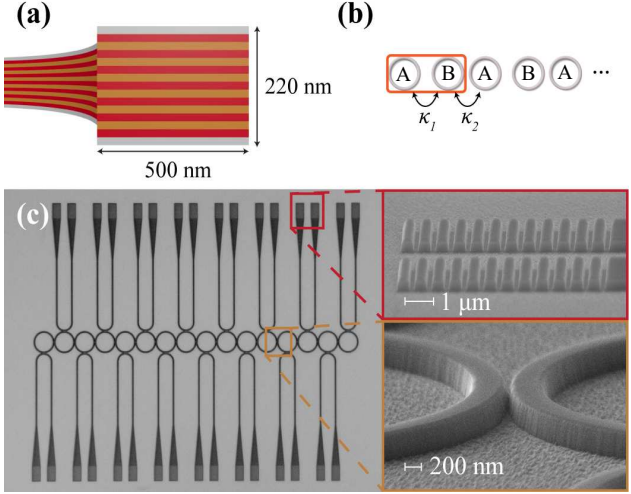}
\caption{(a) InGaAsP multilayer quantum well structure used in the microrings, (b) a schematic of the SSH microring laser array, and (c) a microscope image of the fabricated structure with 16 elements. Insets show scanning electron microscope images of the grating at the end of the out-coupling waveguides, and the coupling region between two microrings.}
\label{Fig1}
\end{figure}
\begin{align}\label{Eq1}
%&i\frac{d\psi_n^A}{dt}=\epsilon_A\psi_n^A+\kappa_1\psi_n^B+\kappa_2\psi_{n-1}^B \nonumber \\
%&i\frac{d\psi_n^A}{dt}=\epsilon_A\psi_n^A+\kappa_1\psi_n^B+\kappa_2\psi_{n-1}^B ,
%\Lambda_m=\frac{\pi k_0 n_0 D^2}{\pi/4+m \pi/2}, \qquad m=1,2,...
&H_0=\epsilon_A\sum_{n}\hat{c}_n^{A\dagger} \hat{c}_n^A + \epsilon_B\sum_{n}\hat{c}_n^{B\dagger} \hat{c}_n^B \nonumber\\
&+ \sum_{n}(\kappa_1(\hat{c}_n^{B\dagger} \hat{c}_n^A + \hat{c}_n^{A\dagger} \hat{c}_n^B) + \kappa_2(\hat{c}_{n-1}^{B\dagger} \hat{c}_n^A + \hat{c}_n^{A\dagger} \hat{c}_{n-1}^B)),
%&+ \sum_{n}(\kappa_1\hat{c}_n^{B\dagger} \hat{c}_n^A + \kappa_2\hat{c}_{n-1}^{B\dagger} \hat{c}_n^A + c.c.),
\end{align}
where $\hat{c}_n^{A\dagger}$ and $\hat{c}_n^{B\dagger}$ denote photon creation operators at site $n$ in the sublattices $A$ and $B$ of this structure, while $\epsilon_A$ and $\epsilon_B$ represent the complex on-site eigenfrequencies (potentials) of the corresponding active rings. Again, $\kappa_1$ and $\kappa_2$ are the intra-cell and inter-cell coupling coefficients, respectively. In momentum space representation, the Bloch mode Hamiltonian can be obtained through a Fourier transform, i.e.:
\begin{equation}\label{Eq2}
H_0(k)=
\begin{pmatrix}
  \epsilon_A & \kappa_1+\kappa_2e^{-ik} \\
  \kappa_1+\kappa_2e^{ik} & \epsilon_B
\end{pmatrix}.
\end{equation}
If the array is Hermitian and is composed of identical rings ($\epsilon_A=\epsilon_B=0$), the Hamiltonian of Eq. (\ref{Eq2}) anti-commutes with the chiral operator $\mathcal{C}=\sigma_z$. Due to the chiral symmetry the eigenvalues are symmetrically distributed around zero, with the two zero-energy edge states being located at the ends of the Brillouin zone $k=\pm\pi$ (Fig. \ref{Fig2} (a)). The field of the edge modes exponentially decays into the bulk (see Fig. \ref{Fig2} (a)). Note that the field distribution of these edge modes is $\pi$-staggered on the sublattices $A$ and $B$. On the other hand, if the SSH structure is active, then the on-site potentials are now purely imaginary, i.e. $\epsilon_A=-ig_A$ and $\epsilon_B=-ig_B$. In this latter case, it is easy to show that the structure no longer possesses $\mathcal{C}$-symmetry, i.e. $\mathcal{C}H_0\mathcal{C}^{-1}\ne -H_0$. In our experimental realization (Fig. \ref{Fig1} (b)), the parameters $g_A$ and $g_B$ are dictated by the linear gain coefficients associated with the two sublattices $A$ and $B$, as induced by differential pumping. In general, the dynamics of such an active SSH lattice are described by the following set of rate equations \cite{Semiconductor_Lasers,Absar_PRA}:
\begin{align}\label{Eq3}
&\frac{dE_n^{A,B}}{dt}=\frac{1}{2}\bigg[-\gamma+\sigma(N_n^{A,B}-1)\bigg](1-i\alpha_H)E_n^{A,B} \nonumber \\
&+ i\kappa_1E_n^{B,A} + i\kappa_2E_{n \mp 1}^{B,A} \nonumber \\
&\frac{dN_n^{A,B}}{dt}=R_{A,B}-N_n^{A,B}/\tau_r-F(N_n^{A,B}-1)|E_n^{A,B}|^2 .
\end{align}
\begin{figure}[h]
\includegraphics[scale=0.4]{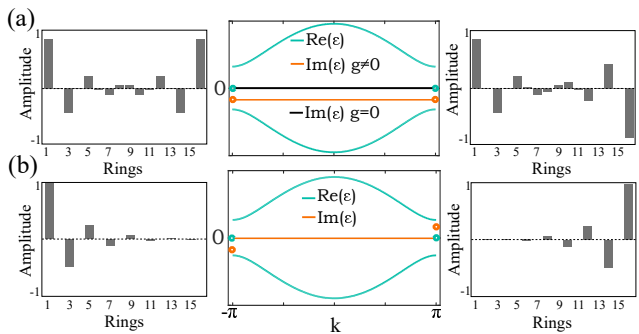}
\caption{Eigenvalue diagrams of (a)Hermitian and uniformly pumped ($g=0$ and $g\neq 0$) and (b) $\mathcal{PT}$−symmetric SSH lattice. The field profiles of the edge states are also depicted in the right and left insets.}
\label{Fig2}
\end{figure}
Here, $E_n^A$ and $E_n^B$ denote the electric modal field amplitudes in sublattices $A$ and $B$, $\gamma$ is inversely proportional to the photon lifetime in each microring cavity, and $N_n^A$ and $N_n^B$ represent the carrier population densities normalized with respect to the transparency value $N_0$. In addition, $\alpha_H$ is the linewidth enhancement factor, $\sigma=\Gamma v_g a N_0$ is proportional to the unsaturated loss in the absence of pumping, $\Gamma$ is the confinement factor, $a$ is the gain proportionality factor, and $v_g=c/n_g$ denotes the group velocity within the cavity. In these equations $R_A$ and $R_B$ are normalized (with respect to $N_0$) pump rates, while $\tau_r$ indicates the carrier recombination lifetime in the InGaAsP quantum wells. Finally, $F=(\Gamma v_g a\epsilon_0 n_e n_g)/2\hbar \omega$, where $\epsilon_0$ is the vacuum permittivity, $n_e$ is the mode effective index, $\hbar$ is the reduced Planck constant, and $\omega$ is the angular frequency of the emitted light. The linear gain coefficients can be directly obtained form the above parameters via $g_{A,B}= (\sigma/2)(R_{A,B}\tau_r-1)-\gamma/2$.

We first consider the simplest possible case, i.e. when the pumping is uniform $g_A=g_B=g$ in this topological arrangement. The linear band structure of this SSH laser system under this condition is also depicted in Fig. \ref{Fig2} (a). The optical field distributions corresponding to the two edge states are identical to those in the Hermitian case. It is evident that in this scenario, the eigenvalues are no longer symmetrically distributed around the zero level, instead, they all shift by the same amount $-ig$ (in their imaginary part), corresponding to an equal amount of gain $g$ for all supermodes involved (red line in the dispersion curve of Fig. \ref{Fig2} (a)). This situation drastically changes once $\mathcal{PT}$-symmetry is introduced, i.e. $g_A=-g_B=g$. In this regime, the Hamiltonian of Eq. (\ref{Eq2}) now takes the form:
\begin{equation}\label{Eq4}
H_0(k)=
\begin{pmatrix}
  -ig & \rho \\
  \rho^* & +ig
\end{pmatrix} ,
\end{equation}
where $\rho=\kappa_1+\kappa_2e^{-ik}$. As previously indicated, this Hamiltonian does not respect the chiral $\mathcal{C}$-symmetry. Instead, $H_0$ satisfies:
\begin{equation}\label{Eq5}
  \mathcal{C}\mathcal{T}_{PS}H_0\mathcal{T}_{PS}^{-1}\mathcal{C}^{-1}=H_0 ,
\end{equation}
where the pseudo-spin time reversal operator $\mathcal{T}_{PS}$ is here defined as $\mathcal{T}_{PS}=i\sigma_y \mathcal{K}$, with $\mathcal{K}$ denoting complex conjugation. If $\ket{\psi}$ is an eigenstate of $H_0$ ($H_0\ket{\psi}=\epsilon\ket{\psi}$), then $\mathcal{C}\mathcal{T}_{PS}\ket{\psi}$ is also an eigenstate of $H_0$, with eigenvalue $\epsilon^*$. While $\mathcal{PT}$-symmetry is imposed in real space, the $\mathcal{CT}$ operator acts in the momentum domain. Figure \ref{Fig2} (b) shows the eigenvalues of this $\mathcal{CT}$-symmetric Hamiltonian when $g<|\kappa_2-\kappa_1|$. The right and left sides of this figure display the field amplitudes of the edge states corresponding to the two imaginary eigenvalues marked in the plot (red dots at the edges of the dispersion diagrams). Evidently, the field distributions of these states only occupy one of the sublattices ($A$ or $B$), and alternate in sign.  As a result, one of these modes is expected to experience gain, while the other one an equal amount of loss. Note that under $\mathcal{PT}$-symmetric conditions, all the bulk modes remain neutral (neither gain nor loss).

\begin{figure}[h]
\includegraphics[scale=0.4]{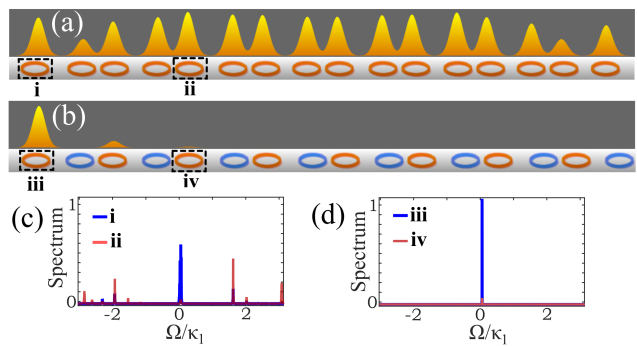}
\caption{Theoretically predicted steady-state lasing profiles for (a) uniformly pumped and (b) $\mathcal{PT}$-symmetric SSH lattice. Panels (c) and (d) depict the power spectra corresponding to the edge mode and bulk modes as obtained from different locations in the array.}
\label{Fig3}
\end{figure}

Figure \ref{Fig3} (a) shows the steady-state intensity distribution as obtained from simulations (Eqs. (\ref{Eq3})), for a 16-element SSH laser system when uniformly pumped at $R_{A,B}=1.06/ \tau_r$. In these simulations, we assume that $\alpha_H=4$, $\tau_r=4 ns$, and $\sigma=6\times 10^{11} s^{-1}$. The dimerization $\nu=\kappa_2/\kappa_1$ for this array is $\nu=2$. The resulting lasing profile is a complex mixture of all the supermodes (including the edge states) supported in this laser array. This is because all modes experience the same gain. Nevertheless, the emission wavelength in the array greatly depends on the site number (Fig. \ref{Fig3} (c)). Our theoretical analysis suggests that the edge state will always lase at the resonance frequency $\omega_0$ ($\Omega/\kappa_1\approx0$) with a relatively narrow linewidth. Conversely, the spectrum emanating from rings in the bulk (bulk modes) will have a considerably more complex structure because of mode competition effects. On the other hand, Fig. \ref{Fig3} (b) shows the expected intensity distribution under $\mathcal{PT}$-symmetric conditions after numerically solving Eqs. (\ref{Eq3}) - starting from noise. In this case, the two sections are pumped at $R_A=1.06/ \tau_r$ while $R_B=1.03/\tau_r$, thus setting sublattice $A$ above lasing threshold, whereas $B$ is kept below threshold. In this regime, our simulations show that only one of the edge modes (the one enjoying gain) is favored and hence lases, while all the bulk modes are suppressed. In direct contrast to the results presented in Fig. \ref{Fig3} (a), $\mathcal{PT}$-symmetry now promotes only the edge state. Consequently, the spectrum emitted from the structure happens to be close to the ring resonance $\omega_0$ and is single-moded (Fig. \ref{Fig3} (d)).

\begin{figure}[h]
\includegraphics[scale=0.4]{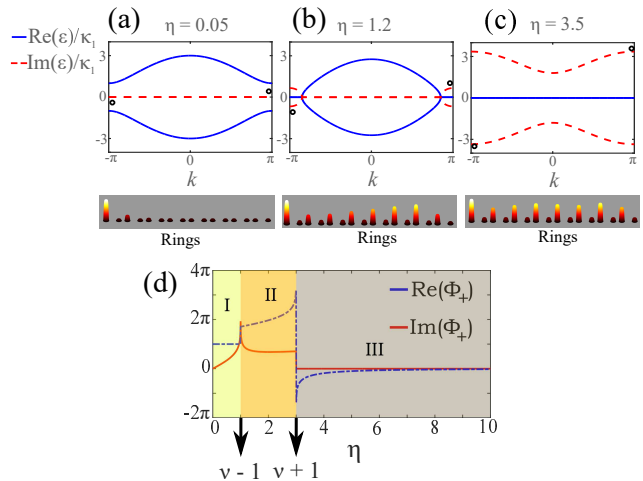}
\caption{Complex band structure of the PT-symmetric SSH model, where (a), (b) and (c) correspond to the three phases I, II and III, respectively. The insets show the simulated intensity distributions corresponding to these three distinct regimes. Panel (d) presents the complex Berry phase $\Phi_+$ as a function of the normalized gain coefficient $\eta$.}
\label{Fig4}
\end{figure}

The dynamics of this $\mathcal{PT}$-symmetric SSH laser system can be theoretically predicted by considering the linear complex band structure associated with the non-Hermitian Hamiltonian of Eq. (\ref{Eq4}), which is given by:
\begin{equation}\label{Eq6}
  \epsilon(k) = \pm \kappa_1 \sqrt{1+\nu^2+2\nu cos(k)- \eta^2} ,
\end{equation}
where $\eta=g/\kappa_1$ represents a normalized gain/loss. This equation reveals three distinct phases, presented in Figs. \ref{Fig4} (a)-(c). If the SSH system is pumped or operated in the range of $0<\eta<\nu-1$ (denoted as phase I), only the edge state is expected to lase. In this domain, under steady-state conditions, the structure is single-moded and the intensity profile across the array varies exponentially with the site number (inset of Fig. \ref{Fig4} (a)). As the gain in the topological system increases, i.e. when $\nu-1<\eta<\nu+1$, the SSH structure enters phase II, where some of the bulk modes start to acquire complex eigenvalues (after entering the $\mathcal{PT}$-symmetry broken phase), resulting in a multimode operation (inset of Fig. \ref{Fig4} (b)). Note that in phase II the intensity profile across the array is asymmetrically one-sided, biased towards the edge mode. Finally, for even higher values of gain/loss contrast, i.e. $\eta>\nu+1$, the array crosses another threshold and moves into phase III, as also corroborated by analyzing Eqs. (\ref{Eq3}). At this point, all of the bulk modes of the active lattice break their $\mathcal{PT}$-symmetry, and as such, they start to lase - all competing for the gain. Unlike what happens in the first two phases, after crossing into phase III, the edge state is now obscured by bulk modes. This in turn results into a more uniform intensity profile, as shown in the inset of Fig. \ref{Fig4} (c). In other words, in this range, the pumped sublattice is uniformly lasing, while its lossy counterpart remains dark. This can be explained by the fact that the carrier-induced detuning between adjacent resonators, which is by nature a nonlinear effect, significantly suppresses the coupling between neighboring units. The theoretically expected spectra corresponding to these three phases can be found in the Supplementary. Our simulations also suggest that the boundary between phase I and II can be nonlinearly modified because of the linewidth enhancement factor $\alpha_H$, something that is also revealed in our experiments. This is analyzed in greater detail in the Supplementary.

Interestingly, the onset of these three phases is also manifested in the complex Berry phase associated with this SSH laser array \cite{Liang_Huang}. Figure \ref{Fig4} (d) shows the Berry phase associated with the upper band $\Phi_+$ as a function of the normalized gain $\eta$ when $\nu=2$. This figure reveals that the geometric phase undergoes phase transitions at exactly the same boundaries ($\nu \pm 1$), as also previously suggested by Eq. (\ref{Eq6}).

\begin{figure}[h]
\includegraphics[scale=0.47]{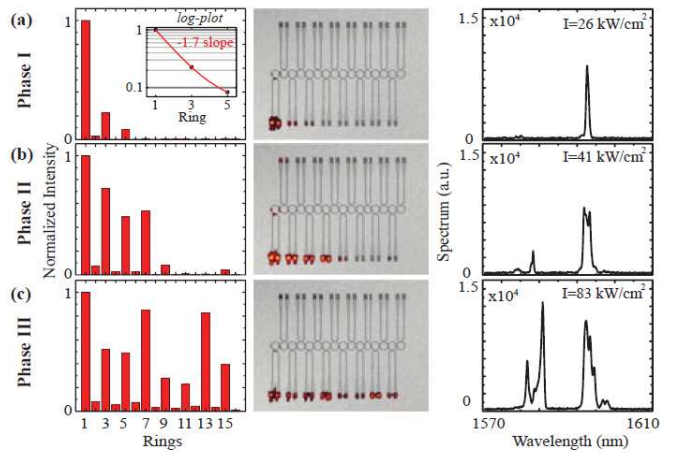}
\caption{The left panels depict the measured intensity distributions in the 16-element SSH array at every site. The middle panels show raw data from the extraction ports, while the right panels the corresponding power spectra. Each of the successive rows (a), (b), and (c) are progressively associated with phase I, II, and III observations. The inset in (a) provides the exponential intensity distribution of the lasing edge-state in a log-linear scale.}
\label{Fig5}
\end{figure}

To verify these predictions we conducted a series of experiments with a 16 microring resonator SSH array, each having a radius of $5 \mu m$.  To enforce single-transverse mode operation at $1.59 \mu m$, the width of the resonators was set to $500 nm$. In order to reduce the lasing threshold, the microrings were surrounded by a low-index dielectric, entailing a higher confinement. As previously indicated, each ring was individually interrogated (intensity-wise and spectrally) through an extraction bus waveguide, featuring a pair of grating out-couplers (Fig. \ref{Fig1} (c)). To introduce $\mathcal{PT}$-symmetry, the microresonators were alternately pumped at $1.06 \mu m$ by using a titanium amplitude mask. Figure \ref{Fig5} shows the measured ring intensity and spectra using an InGaAs camera. In particular, (a), (b), and (c) present the data corresponding to phase I, II, and III. At a pump intensity of $I = 26 kW/cm^2$ only the edge mode lases (Fig. \ref{Fig5} (a)). Once the first phase transition occurs, other modes start competing for the gain (Fig. \ref{Fig5} (b) at $I =41 kW/cm^2$) and eventually the edge mode is obscured (Fig. \ref{Fig5} (c) at $I = 83 kW/cm^2$). The emergence of these three phases is also evident in their spectra. While the spectrum of the edge mode is single-moded, once the first phase transition occurs, bulk modes also appear, with upshifted frequencies as expected from theory (see Supplementary). Figure \ref{Fig5} (a) indicates that the exponential intensity decay of the edge mode (log-linear inset) is in good agreement with theory ($\propto(\frac{\kappa_1}{\kappa_2})^{2n}$) when $\nu=1.7$. In all our experiments we found that gain saturation plays a prominent role in stabilizing the lasing edge mode at different pumping levels. Moreover, the same effects tend to shift the frequency of the bulk modes as a function of the pump power. Yet, the central frequency of the edge mode experiences a negligible shift, thus manifesting its robustness (see Supplementary).

In conclusion, we have observed for the first time lasing of the topological edge modes in an active SSH microring array. Under chiral-time symmetry, the transitions of this non-Hermitian system can be described through both the complex band structure and the corresponding complex Berry phase. The effects of gain saturation and carrier dynamics on the edge-mode properties were systematically investigated. Experimental results obtained from both the spatial and spectral domains were found to be in good agreement with theoretical predictions. Our work can provide new perspectives in understanding some of the fundamental aspects associated with the synergy between non-Hermiticity and topology in active systems.

\begin{acknowledgments}
The authors gratefully acknowledge the financial support from Office of Naval Research (ONR) (N00014-16-1-2640), National Science Foundation (NSF) (ECCS-1454531, DMR-1420620), Air force Office of Scientific Research (AFOSR) (FA9550-14-1-0037), Binational Science Foundation (BSF) (2016381) and Army Research Office (ARO) (W911NF-16-1-0013). This work was also partially funded by the Qatar National Research Fund (NPRP 9-020-1-006).
\end{acknowledgments}

% Create the reference section using BibTeX:
\bibliography{references}
\end{document}